\documentstyle[11pt,aaspp4]{article}

\def\ea{{\it et al.}~}

\def\ie{{\it i.e.,~}}

\newcommand{\be}{\begin{equation}}
\newcommand{\ee}{\end{equation}}

\begin{document}

\title{Ambipolar Diffusion in YSO Jets }

\author{Adam Frank$^1$, Thomas A. Gardiner$^1$, Guy Delemarter$^1$, 
Thibaut Lery$^2$ \& Riccardo Betti$^3$}
\bigskip

\affil{$^1$ Dept. of Physics and Astronomy,\\
University of Rochester, Rochester, NY 14627-0171}
\affil{$^2$ Department of Physics, Queen's University, \\
Kingston, ON K7L 3N6, Canada}
\affil{$^3$ Dept. of Mechanical Engineer, Laboratory for Laser Energetics,\\
University of Rochester, Rochester, NY 14627-0171}
\authoraddr{address used for editorial communication only}


\begin{abstract}

We address the issue of ambipolar diffusion in Herbig-Haro jets.  The
current consensus holds
that 
these jets are launched and collimated via
MHD forces.  Observations have, however, shown that the jets can be
mildly to weakly ionized.  Beginning with a simple model for
cylindrical equilibrium between neutral, plasma and magnetic pressures
we calculate the characteristic time-scale for ambipolar diffusion. Our
results show that a significant fraction of HH jets will have ambipolar
diffusion time-scales equivalent to, or less than the dynamical
time-scales.  This implies that MHD equilibria established at the base
of a HH jet may not be maintained as the jet propagates far from its
source.  For typical jet parameters one finds that the length scale where
ambipolar diffusion should become 
significant
corresponds to the
typical size of large (parsec) scale jets. We discuss the significance
of these results for the issue of magnetic fields in parsec-scale
jets.

\end{abstract}


\section{INTRODUCTION}

Narrow hypersonic jets are a ubiquitous phenomena associated with star
formation. In spite of the large database of multi-wavelength
observations and numerous theoretical studies, a number of fundamental
questions remain unanswered about these jets.  Paramount among the
outstanding issues is the role of magnetic fields in the launching,
collimation and propagation of Young Stellar Object (YSO) jets.  The
current consensus holds that the jets are formed on small scales ( $L
< 10 ~AU$) via magneto-centrifugal processes associated with accretion
disks (\cite{Burea96}, \cite{Shuea94}, \cite{Ouyed1997a}).  At larger
scales, where jet propagation rather than collimation is the issue,
there is an implicit assumption that the magnetic fields involved in
the launching process will remain embedded in the jets as they
traverse the inter-cloud medium.
Observations of narrowly collimated jets, or chains of bow-shocks,
extending out to parsec scale distances (\cite{Reipurth1997}) has
strengthened the viewpoint that the jet beams must ``carry their own
collimators''.  This is based on the possibility 
that without dynamically significant magnetic fields confining the
beam, the jets may be disrupted by hydrodynamical instabilities
(\cite{Stoneea97}, \cite{Hardeea97}).

Do the magnetic fields remain embedded in the jets?  That is the
question we address in this paper.  If we accept that jets are created
via MHD forces then the only way to lose the imposed fields is via
reconnection or through diffusive processes.  The helical topology
expected for magneto-centrifugally launched jets would not be likely to lead
to large scale reconnection throughout the beam (though reconnection
may have important effects on the radiative properties of the jets,
\cite{Gardinea99}).  Thus, if magnetic fields can be cleared out of a
jet this is more likely to occur via diffusive processes.  Large format
CCD mosaics have recently demonstrated that YSO jets can extend over
multi-parsec length scales ($L_j \approx 3$ pc).  The dynamical age
($t_{dyn} = L_j/V_j$) for these jets falls in the range $t_{dyn} =
10^4 ~ - ~ 10^5$ y (\cite{Reipurth1997}, \cite{EisMun1997}). Thus,
even if diffusive processes are slow, they will have a relatively long
time to affect
the propagation of the beams.  In plasmas with less than full
ionization, ambipolar diffusion is one of the most effective means of
driving the rearrangement of flux relative to the plasma.
Observations of optical Herbig-Haro (HH) jets have shown them to be
moderately ionized, with ionization fractions ranging from $x =
.5~-~.01$ (\cite{Hartiganea96}, \cite{BacEis99}), where the ionization
fraction is defined as $x = n_i/n_n$ and $n_n$ and $n_i$ are the
number densities of neutral and ionized particles respectively. 
The ionization fractions are seen to fall as the plasma 
propagates away from internal shocks and may, therefore, fall to even
lower values in the optically invisible regions of the beam.  Given
the potentially low ionization fraction and the extreme ratio of jet
radius to length, $R_j/L_j < 10^{-3}$, it may be possible for
ambipolar diffusion to significantly alter the 
MHD equilibria established when the jet was launched.

In this paper we wish to address the issue of ambipolar diffusion in
YSO jets.  Our present goal is simply to establish the characteristic
time and length scales for ambipolar diffusion to operate with respect to the
fundamental jet parameters.  We leave its consequences to later
papers.  In section 2 we establish the conditions for an MHD
equilibrium in a simplified model of a jet in terms of plasma, neutral
and magnetic pressures.  In section 3 we consider the equation for
ion-neutral drift in this context and in section 4 we present an
equation for the ambipolar diffusion time-scale in YSO jets.  In
section 5 we discuss our results in light of observations and other
models.

\section{Initial Configuration}

We begin by considering the simplest model of a magnetized jet.  We
imagine a cylindrical column of material of radius $R_j$.  The column
is composed of ions, electrons and neutral species and is initially
held together internally by a radial balance of  pressure ($P_i, P_e,
P_n$) and magnetic ($\vec{J} \times \vec{B})$ forces. We further
assume that the column is in force balance with a (possibly
magnetized) ambient medium.  Thus, initially, the jet does not expand
laterally (\ie in the radial direction).  To simplify our calculation
we assume that the temperature of all species is the same ($T_i = T_e =
T_n$).  We further assume that the magnetic field in the jet is purely
poloidal, $\vec{B} = B_z(r) \hat{e}_z$.  Figure 1 shows a schematic of our
model for the magnetized jet.

\subsection{Global Force Balance}

Our calculation is similar to derivations of the ambipolar diffusion
time-scale in collapsing molecular clouds (a description of those
models can be found in \cite{Spitzer78} and \cite{Mouschovias91}).
Molecular clouds are cold gravitationally bound systems of extremely
low ionization ($ x \approx 10^{-6} $).  This implies that terms
involving gas pressure and ion density can be dropped from
consideration when calculating the effects of ambipolar
diffusion. Gravity does not, however, play a role in maintaining the
cross-sectional properties of a YSO jet and pressure forces can not be
ignored. In addition, given the range of ionization fractions in these
systems the ion inertia should not, {\it a priori}, be dropped from
calculations. In what follows we derive a simple expression for the
ambipolar diffusion time-scale (based on the ion-neutral drift velocity)
including the effects of gas and plasma pressure as well as the ion
density.

We begin with the three fluid force balance equations for electrons
($e$), ions ($i$) and neutrals ($n$) which can be written as
\begin{eqnarray}
\rho_e {d \vec{v}_e \over dt} & = & - \vec{\nabla} P_e - n_e e \left(
\vec{E} + {\vec{v}_e \over c} \times \vec{B} \right)
 + \vec{F}_{ei} + \vec{F}_{en}, \\
\rho_i {d \vec{v}_i \over dt} & = & - \vec{\nabla} P_i + n_i Z e \left(
\vec{E} + {\vec{v}_i \over c} \times \vec{B} \right)
 + \vec{F}_{ie} + \vec{F}_{in}, \\
\rho_n {d \vec{v}_n \over dt} & = & - \vec{\nabla} P_n 
 + \vec{F}_{ne} + \vec{F}_{ni},
\end{eqnarray}
where the term $d/dt$ represents the convective derivative and
$F_{12}$ represents the drag force imparted on species 1 by 
colliding with species 2.  Note that $F_{12} = - F_{21}$.

In what follows we will ignore the electron inertia term.  This is
valid when the time-scales of interest are long compared with the
response time of the electrons.  Formally this means that the
dynamical time $t_{dyn}$ is longer than either the electron cyclotron
period $2\pi/\Omega_{ec}$ or the electron plasma period
$2\pi/\omega_{ep}$ (\cite{elliot93}) as is the case in YSO jets with
fields greater than the $\mu G$ level.  We will also neglect the
electron collisional coupling to the the neutrals, $F_{en}$, given the
difference in masses, $M_e << M_i,M_n$ (see footnote \#4
\cite{Mous1996}, Ciolek private communication 1999).

With these caveats the addition of equations 1 and 2 yields the following
expression:
\begin{equation}
\rho_i {d \vec{v}_i \over dt}= - \vec{\nabla} (P_i + P_e) +
(n_i Z e - n_e e) \vec{E} + 
\left( {n_i Z e \vec{v}_i - n_e e \vec{v}_e \over c} \times \vec{B} \right) + 
\vec{F}_{in}. \label{add12}
\end{equation}

\noindent Assuming charge neutrality ($n_i Z  = n_e$) yields the
following form for the current density,
\begin{equation} 
\vec{J} =  n_i Z e \vec{v}_i - n_e e \vec{v}_e = n_e e
(\vec{v}_i - \vec{v}_e), \label{currdef}
\end{equation}

\noindent and the term with the electric field drops out of equation 
\ref{add12}.  Making use of Amp\'ere's law, the Lorentz force can be 
decomposed into two terms,
\begin{equation} {1 \over c}(\vec{J} \times \vec{B}) = -
\vec{\nabla}_\perp {B^2 \over 8 \pi} + {B^2 \over 4 \pi R_c} \hat{n}
\label{lforce}
\end{equation}

\noindent 
where $\hat{n}$ is a unit vector directed toward the local center of 
curvature of the field line, $R_c$ is the local radius of curvature, and 
$\vec{\nabla}_{\perp}$ is the gradient normal to field lines.

Our choice of a $\vec{B} = B_z(r) \hat{e}_z$ implies that the second
term in equation \ref{lforce} is zero.
We are left, therefore, with only the first term which can be identified as
the gradient of magnetic pressure, $P_B = B^2/8\pi$.

Finally, if we add the momentum equations for all three species, again
ignoring the electron inertia we arrive at the following equation,
\begin{equation}
\rho_i  {d \vec{v}_i \over dt} + \rho_n {d \vec{v}_n
\over dt} = - \vec{\nabla} (P_i + P_e + P_n + P_B). \label{forcebal} 
\end{equation}

\noindent In what follows we need only consider the radial component 
of the force equation.  Thus if,
\begin{equation} 
{\partial \over \partial r} (P_i  + P_e + P_n + P_B) = 0, \label{Gpbal} 
\end{equation}

\noindent and ${v}_{i,r} = {v}_{n,r} = {v}_{e,r} = 0$ (where the $v_r$'s 
refer to radial velocities) at $t = 0$ then the entire configuration
begins in equilibrium.  It is, however, what we shall call a {\it
quasi-equilibrium} since each species will be accelerated separately
by the terms on the RHS of equations 1 - 3. The plasma and the
neutrals will attempt to move past each other as each responds to its
own forces. It is only the collisional drag $F_{ij}$ between species
which holds the initial configuration together. This is exactly the
situation which occurs in a molecular cloud where the cloud can be
supported for some period of time by its magnetic field.  As in the
molecular cloud case, the quasi-equilibrium in the jet can not persist
indefinitely and the initial configuration will change on a time-scale
$t_{ad} = R_j/v_d$ where $v_d$ is the ion-neutral drift velocity $v_d
= v_n - v_i$. We seek to calculate this time-scale.

Note that equation \ref{Gpbal} implies that the pressure distributions
of the ions and neutrals is not independent.
\begin{equation} 
{\partial \over \partial r} P_n = -
{\partial \over \partial r} (P_i + P_e + P_B) \label{pndep} 
\end{equation}

\noindent We will use this relation below in our calculation of $t_{ad}$.

\section{The Ion-Neutral Drift}

If we divide equation \ref{add12} by $\rho_i$ and subtract from it
equation 3 (itself divided by $\rho_n$) we find,
\begin{equation} 
{d \vec{v}_i \over dt} - {d \vec{v}_n \over dt} = -
{1 \over \rho_i} \vec{\nabla} (P_i + P_e + P_B) + {1 \over \rho_n} 
\vec{\nabla} P_n
+ \vec{F}_{in} \left({\rho_n + \rho_i \over \rho_n \rho_i}\right).
\label{that}
\end{equation}

\noindent If we now assume that the collisions will quickly bring the
ion-neutral drift to a steady velocity, \ie the relative
accelerations approach zero after a few collisions, we have the
following radial balance equation between collisions and
hydro-magnetic forces,
\begin{equation}
F_{in}\left({\rho_n + \rho_i \over \rho_n \rho_i}\right) = 
{1 \over \rho_i} {\partial \over \partial r} (P_i + P_e + P_B) - 
{1 \over \rho_n} {\partial \over \partial r} P_n. \label{driftbal}
\end{equation}

\noindent Substituting equation \ref{pndep} into the above expression
we find the densities cancel, leading to
\begin{equation} 
{F}_{in} = {\partial \over \partial r}  (P_i + P_e + P_B). \label{fcon1} 
\end{equation}

\noindent  We now make the approximation that the scale of the gradients 
in the jet are such that
\begin{equation}
{\partial P_k \over \partial r} \approx {P_{k} \over R_j}
\label{pscale}
\end{equation}

\noindent where $P_k$ is is the characteristic
scale for that pressure of the k-th component of the configuration.
We also use the definition of the plasma $\beta$ parameter, $\beta =
P_g/P_B$, where $P_g = P_i + P_e$.  Thus
\begin{eqnarray}
F_{in} & = & {P_{B} \over R_j} (\beta +1) \\
       & = & { {B}^2 \over 8 \pi R_j} (\beta +1).
\end{eqnarray}

\noindent To reduce this equation further we must consider the form of 
the collisional force.  $F_{in}$ can be written as,
\begin{equation} 
\vec{F}_{in} = \rho_i \rho_n {\langle\sigma w \rangle_{in} \over
M_i + M_n} (\vec{v_n} - \vec{v_i}), \label{fdef} 
\end{equation}

\noindent
where $\langle\sigma w \rangle_{in} \approx 3 \times 10^{-9} ~cm^3 ~s^{-1}$ 
is the average collision rate (\cite{Spitzer78}).  We can now solve for the 
ion-neutral drift velocity.
\begin{equation} 
v_d = {M_i + M_n \over 8 \pi} {1 \over  \langle\sigma w \rangle_{in} }
\left( {B^2 \over \rho_i \rho_n R_j} \right) (\beta + 1). \label{vdeq} 
\end{equation}

\noindent 
Thus the time-scale for changes in the initial configuration 
can be written as
\begin{equation} 
t_{ad} = {8 \pi \langle\sigma w \rangle_{in} M_n M_i \over M_i + M_n} 
x \left({n _n R_j \over B} \right)^2 ( \beta + 1)^{-1},  \label{vdeq1}
\end{equation}

\noindent where we have used $x = n_i/n_n$. The above expression gives the
ambipolar diffusion time-scale in terms of the fundamental parameters in the
jet ($x, n_n, R_j$ and $B$).  

\section{Ambipolar Diffusion in YSO Jets}

Of the four variables in equation \ref{vdeq1} only first three ($x$,
$n_n$, $R_j$) are well characterized for HH jets.  From observations
typical values are: $.01 \le x \le .1$; $10^3 \le n_n/~cm^{-3} \le
10^4$; $1 \times 10^{15}~ \le R_j/~cm \le 5 \times 10^{15}$
(\cite{BacEis99}).  Magnetic fields however are not so easily
categorized.  To date there have been only a handful of measurements
of fields in YSO jets (\cite{Ray1997}).  Thus, it would be better to
cast equation \ref{vdeq1} in terms of a parameter such as the
temperature which has been measured in many YSO jets.  Typical values
in optical jets are $5 \times 10^3 ~K < T < 3 \times 10^4 ~K$,
(\cite{BacEis99}).  To express equation \ref{vdeq1} in terms of T we
use our assumption that the electron and ion temperature is the same
and, furthermore, assume that the plasma is mainly composed of
Hydrogen atoms. Thus
\begin{eqnarray}
M_i & = & M_n  \\
n_i  & = & n_e   \\
P_e = n_e k T & = & n_i k T = P_i
\end{eqnarray}

\noindent The magnetic field can now be expressed as
\begin{equation}
B^2 = 8\pi \frac{P_e + P_i}{\beta} = 16 \pi n_i k T {1 \over \beta},
\end{equation}

\noindent and the ambipolar diffusion time-scale becomes
\begin{equation} 
t_{ad} = {M_n \langle\sigma w \rangle_{in} \over 4 k } 
\left( { n_n R_j^2 \over T_j} \right) 
\left( {\beta \over \beta + 1} \right).
\label{vdeq2} \end{equation}

\noindent Upon normalization the ambipolar diffusion time-scale can be 
written as
\begin{eqnarray} 
t_{ad} & = & 28,904
\left( { n_n \over 10^3 cm^{-3}} \right) 
\left( { R_j \over 10^{15} cm} \right)^2 
\left( { 10^4 K \over T_j} \right) Q(\beta)  ~ y  \\
Q(\beta) & = & \left( {\beta \over \beta + 1} \right)
\label{vdeq3} 
\end{eqnarray}

\noindent Note that $Q(\beta)$ is positive with an asymptotic value 
of $1$ as $\beta \rightarrow \infty$.

This shows that the ion-neutral drift can be driven by gas pressure
forces alone if the individual species' pressure gradients have
different signs, an artificial situation not likely to be encountered
in real jets.  Detailed calculations of equilibrium magnetically
confined jets arising from accretion disks indicate that $\beta$ can
become quite low in some regions of the jet.   
As an example, we provide in Figure 2 a plot of $\beta$ versus radius
for an equilibrium MHD jet. This model was calculated via the Given
Inner Geometry method of Lery \ea (\cite{Leryea98}, \cite{Leryea99})
for determining the asymptotic structure of magneto-centrifugally
launched flows.  Figure 2 shows that $\beta$ can drop to values below
$\beta = 1$ in the {\it inner} regions of the jet ($r < R_j$).

In Figure 3 and 4 we show surface plots of $t_{ad} ~vs~
\beta$ for $\beta = 1$ and $\beta = .1$.  From equations 25 and 26 and
these figures it is clear that depending on the conditions in the
jets, $t_{ad}$ can become as large as $10^6$ y and as small as $10^3$
y.  For jet parameters in the middle of the expected range of
variation we find $t_{ad}$ of order $10^5$ to $10^4$ y.  What is
noteworthy about these results is a significant region of parameter
space exists where the dynamical time-scale for YSO jets is of order
of, or greater than, the ambipolar diffusion time: $t_{dyn} > t_{ad}$.
{\it Thus ambipolar diffusion is likely to play a role in the dynamics
of large scale YSO jets and outflows}.  We discuss the consequences of
this conclusion in the next section.

\section{Discussion and Conclusions}

Our conclusion that ambipolar diffusion time-scales can be comparable
to jet dynamical lifetimes raises a number of intriguing issues.  The
first is the most obvious.  Ambipolar diffusion will rearrange the
mass to flux ratio in the jet and alter any initial equilibrium
between the plasma, neutrals and magnetic field.  Our analysis is too
simplified to yield conclusions about how the jet will evolve in the
presence of ambipolar diffusion.  In our analysis we did not consider
the effect of toroidal fields.  The hoop stresses associated with
toroidal fields will the pull ions towards the jet axis.  
Thus depending on the orientation of
the magnetic pressure gradients, magnetic forces (pressure and tension)
can either compete or apply forces in the same direction.  Including a
toroidal component does not, in general, change the order of
magnitude of the ambipolar diffusion time-scale, however, it can
change the direction in which the ions are pushed. The plasma and
magnetic field may bleed out of the jet leaving only the neutrals or,
conversely, strong hoop stresses could draw the plasma and field in
towards the jet axis.  In both cases, however, the neutrals will be
free to contract or expand depending on their pressure distribution
relative to the ambient medium.  Thus one expects a potentially
significant rearrangement of the jet's cross sectional properties when
ambipolar and jet time-scales are comparable.  The consequences of
ambipolar diffusion on long term jet propagation should, therefore, be
investigated in detail.

The possibility that many jets will have $t_{dyn} \approx t_{ad}$ is
suggestive for the general issue of YSO outflows. There remains
considerable debate over the connection between HH jets and molecular
outflows (\cite{MasCher93}, \cite{Cabrit97}).  The discovery of parsec
scale ``superjets'' (\cite{Reipurth1997}, \cite{EisMun1997}) has
strengthened the case for jets as the source of molecular outflows by
extending jet lifetimes.  Still, most outflows have larger dynamical timescales 
than even the superjets by a factor of at least a few.  One may wonder then why the
jets have shorter lifetimes then the molecular outflows.  Our results
suggest that in some systems at least it is ambipolar diffusion which
sets the lifetime of the visible collimated jet.  Converting to
consideration of length scales, if one takes the minimum ambipolar
diffusion time from our calculation, $t_{ad}\approx 10^3 ~y$ and the
minimum characteristic jet velocity, $V_j \approx 10^2 ~km/s$ one finds
the minimum distance where ambipolar diffusion becomes effective:
$D_{ad} \approx 0.3 ~pc$  Our results indicate that ambipolar diffusion
should not be effective before a jet reaches this distance.  Moreover
for typical values $t_{ad} \approx 10^4 ~y$ and $V_j \approx 3 \times 10^2
~km/s$ one finds $D_{ad} \approx 3 ~pc$. It is indeed noteworthy that
this value corresponds well with the distance where the longest jets
start to fade away.  When $t_{dyn}$ becomes comparable to $t_{ad}$ the
beam may no longer be confined by magnetic forces. 
\acknowledgements

We wish to thank Glenn Ciolek, Larry Helfer and Al Simon
for their help with this project.  We also acknowledge the helpful
comments of Tom Ray, Francesca Bacciotti and Alex Raga.
This work was supported by NSF Grant
AST-0978765


\begin{figure}
\epsfxsize=2.0in
\centerline{\epsfbox{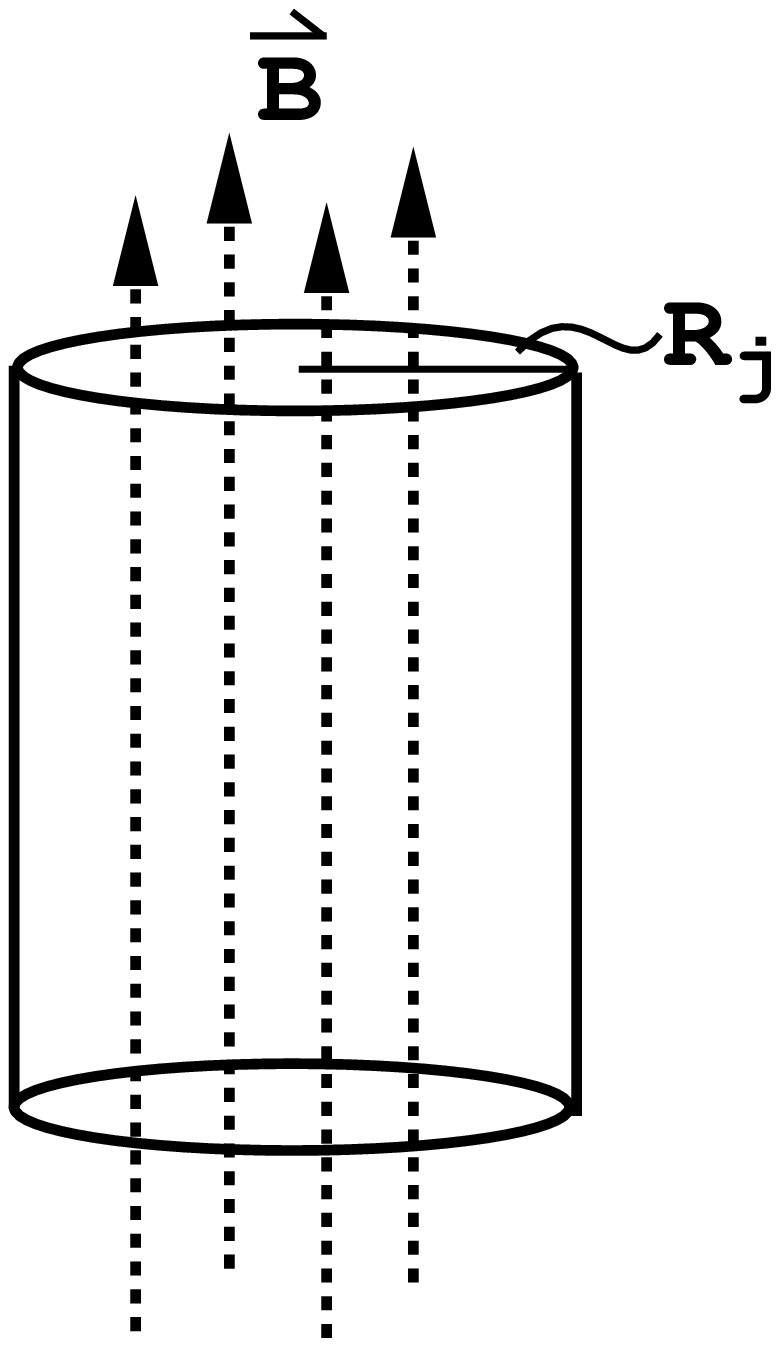}}
\figcaption{Cartoon of Jet Model. Shown is a section of a magnetized jet
composed of a plasma, neutral gas and magnetic field oriented along
the axis of the jet.  The pressure and field distributions are a function
of cylindrical radius only.}
\end{figure}

\begin{figure}
\epsfxsize=6.0in
\centerline{\epsfbox{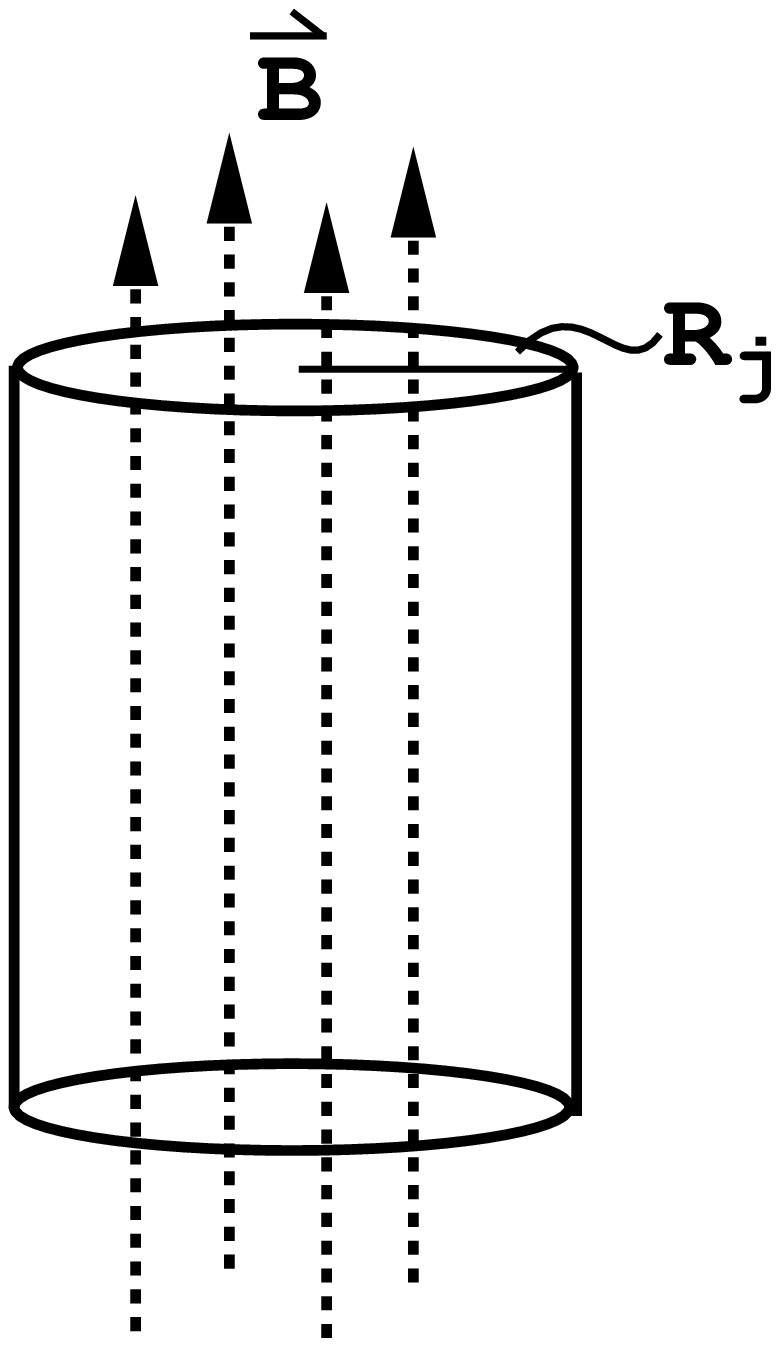}}
\figcaption{Plasma beta vs jet radius for MHD equilibrium jet.
Jet model taken from calculations of Lery \ea 1999.  Note that lowest values
of beta occur within}
\end{figure}

\begin{figure}
\epsfxsize=6.0in
\centerline{\epsfbox{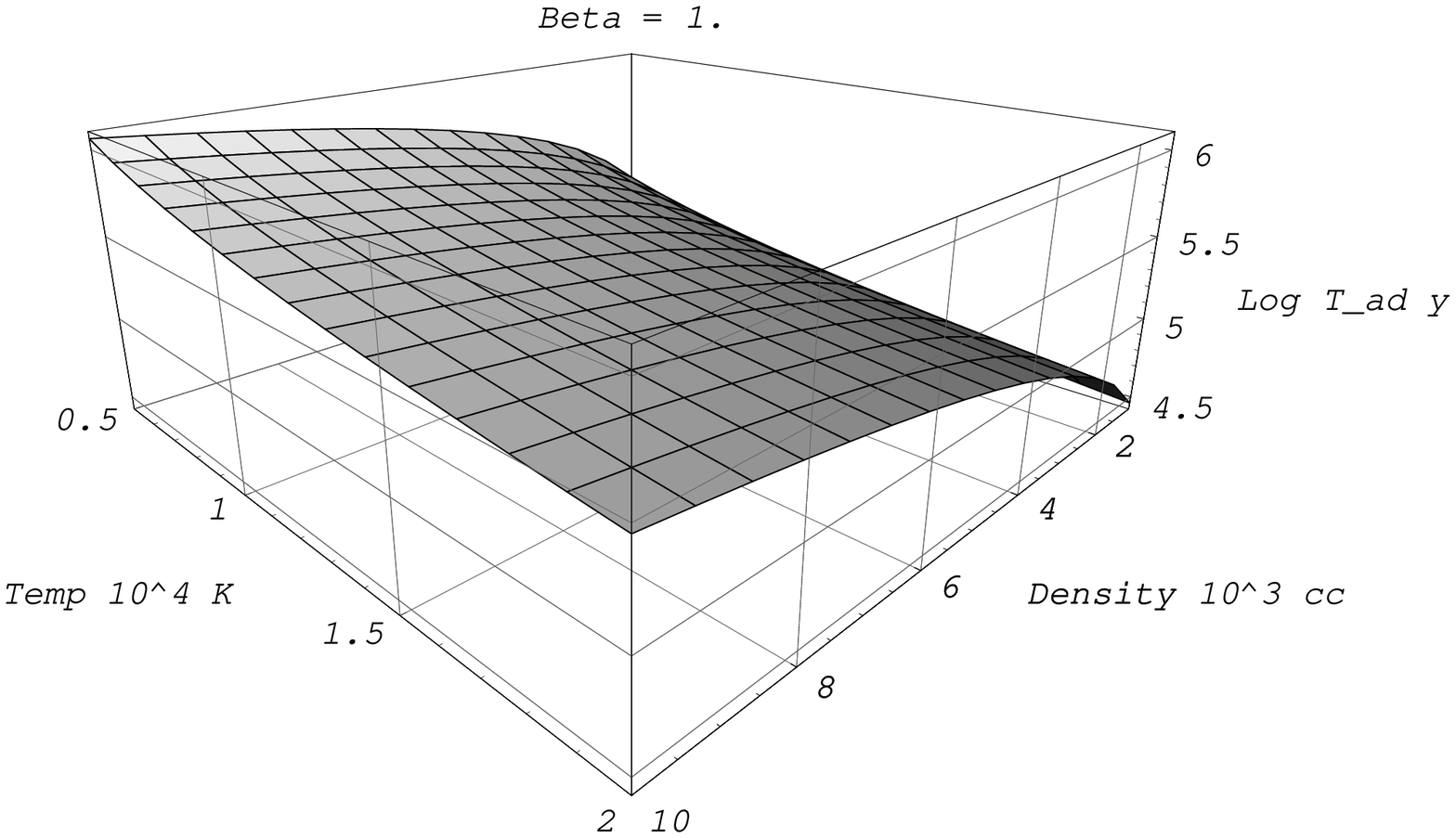}}
\figcaption{The Ambipolar diffusion time-scale as a function of
temperature (in units of 10,000 K) and density (in units of 
1000 $cm^{-3}$) when $\beta = 1$.  The jet radius is $2 \times 10^{15}$ cm.}
\end{figure}

\begin{figure}
\epsfxsize=6.0in
\centerline{\epsfbox{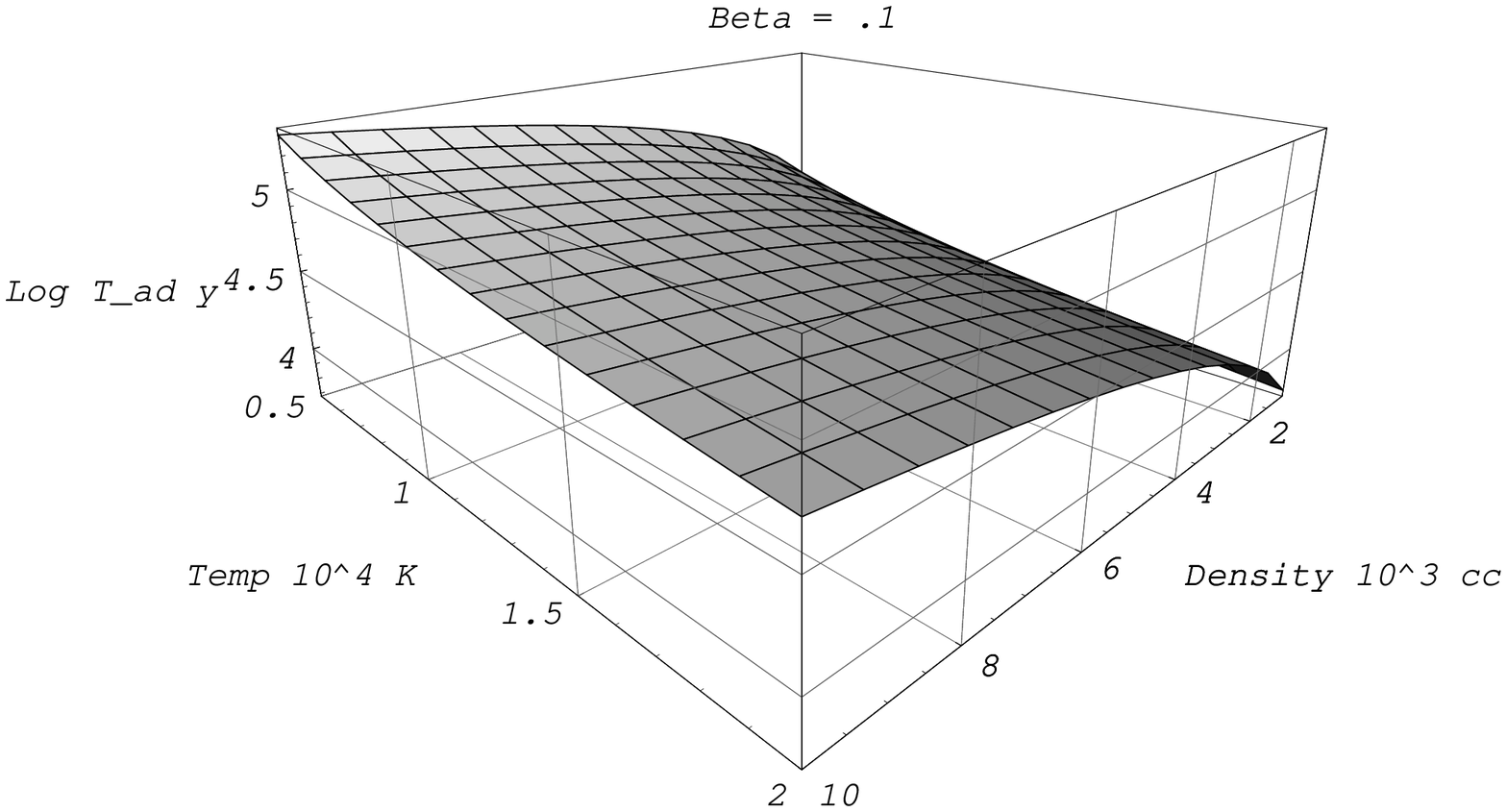}}
\figcaption{The Ambipolar diffusion time-scale as a function of
temperature (in units of 10,000 K) and density (in units of 
1000 $cm^{-3}$) when $\beta = .1$.  The jet radius is $2 \times 10^{15}$ cm.}
\end{figure}

\end{document}